\PassOptionsToPackage{hyphens}{url}
\documentclass[oneside,10pt]{article}

\usepackage[utf8]{inputenc}
\usepackage[slovak,english]{babel}

\usepackage{graphicx}
\usepackage{biblatex}
\usepackage[dvipsnames]{xcolor} 
\usepackage{amsmath, amsthm, amssymb}
\usepackage{mathtools}
\usepackage{hyperref}
\usepackage{siunitx,booktabs,array} 
\usepackage{subfig}
\usepackage[toc,page]{appendix}

\usepackage[bottom]{footmisc}
\usepackage[total={14cm,25cm}, top=20mm, bottom=20mm, right=40mm, left=40mm, includefoot]{geometry}

\usepackage{tikz}

\title{Geometric Illumination of Implicit Surfaces}

\author{
\parbox{0.45\textwidth}{\centering
Michal Zamboj\\[1mm]
Charles University\\
Faculty of Education\\
Department of Mathematics and
Mathematical Education\\
M. Rettigové 4, 116 39\\
Prague, Czech Republic\\[1mm]
michal.zamboj@pedf.cuni.cz
}  
\hspace{0.05\textwidth}
\parbox{0.45\textwidth}{\centering
Jakub {\v{R}}ada\\[1mm]
Charles University\\
Faculty of Mathematics and Physics\\
Mathematical Institute\\
Sokolovská 49/83, 186 00 \\
Prague, Czech Republic\\
~\\[1mm]
rada@karlin.mff.cuni.cz
}
}

\usepackage[hyphens]{url}
\usepackage[hyphenbreaks]{breakurl}

\usepackage{pgfplots}
\pgfplotsset{compat=1.15}
\usepackage{mathrsfs}
\usetikzlibrary{arrows}

\usepackage{xcolor}

\begin{document}

\maketitle
\begin{abstract}
\noindent
Illumination of scenes is usually generated in computer graphics using polygonal meshes. In this paper, we present a geometric method using projections. Starting from an implicit polynomial equation of a surface in 3-D or a curve in 2-D, we provide a semi-algebraic representation of each part of the construction. To solve polynomial condition systems and find constrained regions, we apply algebraic computational algorithms for computing the Gr{\" o}bner basis and cylindrical algebraic decomposition. The final selection of illuminated and self-shaded components for polynomial surfaces of a degree higher than three is discussed. The text is accompanied by visualizations of illumination of surfaces up to degree eight. 
\end{abstract}
\subsection*{Keywords}
illumination, implicit surface, polarity, shadow, cylindrical algebraic decomposition

\vspace*{1.0\baselineskip}
\section{Introduction}
Visual understanding plays a vital role in geometry. Adding a light source to an artificial or natural geometric scene literally sheds light on the geometric properties and displacement of included objects. Creating visualizations with shadows is usually done using polygonal meshes in computer graphics. Such methods are undoubtedly useful for realistic scenes with multiple objects. However, they bring new singularities and special cases into the process.
In contrast, we describe a method based on algebraic (and semi-algebraic) representations until the final plotting by CAS (computer algebra system). In this manner, a system of polynomial equations or inequalities describes each curve, surface, or region in the scene. The backbone of our solution to construct shadows from a point light source is finding a line of tangency (terminator line) and projecting it to all surfaces in the scene. In the second step, we have to select illuminated, shaded, and self-shaded regions of the scene. The presented method is convenient for geometric examination or other mathematical visualizations, since we do not lose valuable information about the given shapes. The drawback of the process is the computational complexity of elimination in polynomial systems. 

\subsection{Literature overview}

For an introduction to applying algebraic concepts to the study of curves and surfaces, see, e.g., the textbooks \cite{Bydzovsky1948,Casas}. Solving polynomial systems using the Gr{\" o}bner basis and the Dixon resultant is discussed in \cite{Kapur1992,Cox2015}. In this paper, we used the implementations in {\sl Wolfram Mathematica 14} \cite{Wolfram1991,Lichtblau2023}. An application of computational methods (in software {\sl Fermat}) to find intersections of surfaces is discussed in various examples in \cite{Lewis2018}. The inverse problem of reconstructing algebraic surfaces from given occlusion contours is treated in \cite{Kang2001}. Computer graphics solutions to create realistic shadows are described, for example, in \cite{Assarsson2003,Stich2007}. An alternative construction of occluding contours through polygonal meshes is described in \cite{Liu2023}. Our implicit representation approach was implemented in connected research on lighting four-dimensional scenes \cite{Zamboj2024}. In this paper, we explain the details of the illumination algorithm in several 2-D and 3-D examples.  

\section{Preliminaries}
This section briefly summarizes the necessary concepts, definitions, and notation.

We work over surfaces (and curves) given by polynomial equations in rational coefficients. If $\mathcal{S}$ is a surface, we denote its polynomial $\sigma$, and similarly in a 2-D case for a curve $c$ and its polynomial $\gamma$. Substitution of the coordinates of a point $P$ into the polynomial $\sigma$ is denoted $\sigma(P)$.

It will be convenient to represent objects in projective homogeneous coordinates to find polars. Let us have $(x'_1,x'_2,\dots,x'_n,x'_0)$, $(y'_1,y'_2,\dots,y'_n,y'_0)\in$ $\mathbb{R}^{n+1}\setminus\{(0,0,\dots,0)\}$. We define the equivalence $$(x'_1,x'_2,\dots,x'_n,x'_0)\sim(y'_1,y'_2,\dots,y'_n,y'_0),$$ if there exists $\lambda\in\mathbb{R}\setminus\{0\}$ such that $(x'_1,x'_2,\dots,x'_n,x'_0)=(\lambda y'_1,\lambda y'_2,\dots,\lambda y'_n,\lambda y'_0)$. The equivalence classes of $\mathbb{R}^{n+1}\setminus\{(0,0,\dots,0)\}$ are called a projective $n$-space. A~point $\overline{P}$ in the projective space has its projective homogeneous coordinates $(p'_1,p'_2,\dots,p'_n,p'_0)$. The transition between projective and Cartesian coordinates is given by substituting $P[\frac{p'_1}{p'_0},\frac{p'_2}{p'_0},\dots, \frac{p'_n}{p'_0}]$, for $p'_0\neq 0$. A point $P$ or polynomial $\alpha$ in Cartesian coordinates will be denoted as $\overline{P}$ or $\overline{\alpha}$ in homogeneous coordinates.

Let $\overline{P}$ be a regular point on a surface $\mathcal{S}$ and assume a polynomial $\overline{\sigma_P}=\overline{P}^T\nabla \overline{\sigma}$. A polar surface $\mathcal{S}_P: \overline{\sigma_P}=0$ is called the first polar of the point $\overline{P}$ with respect to the surface $\mathcal{S}$. Similarly, we define the first polars (or polar lines) for curves.

Throughout the construction, we use two essential computational algorithms. First, to find implicit equations for tangent cones, we need a method to eliminate variables of a polynomial system. For such purposes, we will use the Gr{\" o}bner basis (GB). A proper definition of Gr{\" o}bner basis of an ideal of a set of polynomials requires more extensive preparation and depends on the ordering of monomials (see \cite{Cox2015} for details). Simplifying, the Gr{\" o}bner basis will give us polynomials with the same set of solutions as the original system, but in a reduced form. Alternatively, we can use other elimination methods, such as the Dixon resultant (see \cite{Kapur1992} for an introduction and \cite{Lewis2018,Zamboj2024} for applications). The second algorithm is a cylindrical algebraic decomposition (CAD) of $\mathbb{R}^{n}$ into semi-algebraic subsets separated by polynomial conditions (see \cite{Basu2006} for details). We use the algorithm to partition different regions of a surface included in one conical subspace. Although both algorithms are very powerful tools, they have doubly exponential computational complexity.  

\section{Constructing shadows}
The steps of the illumination algorithm are explained simultaneously in 2-D and 3-D.

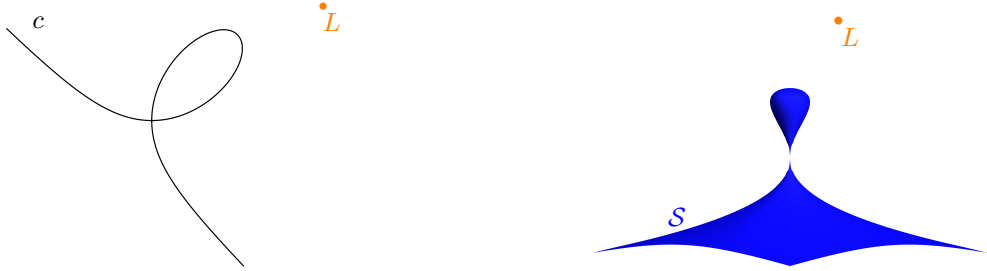
\begin{figure}[!htb]
\centering
	\subfloat{{
 \begin{tikzpicture}
    \node at (0,0){ \includegraphics[height=100pt]{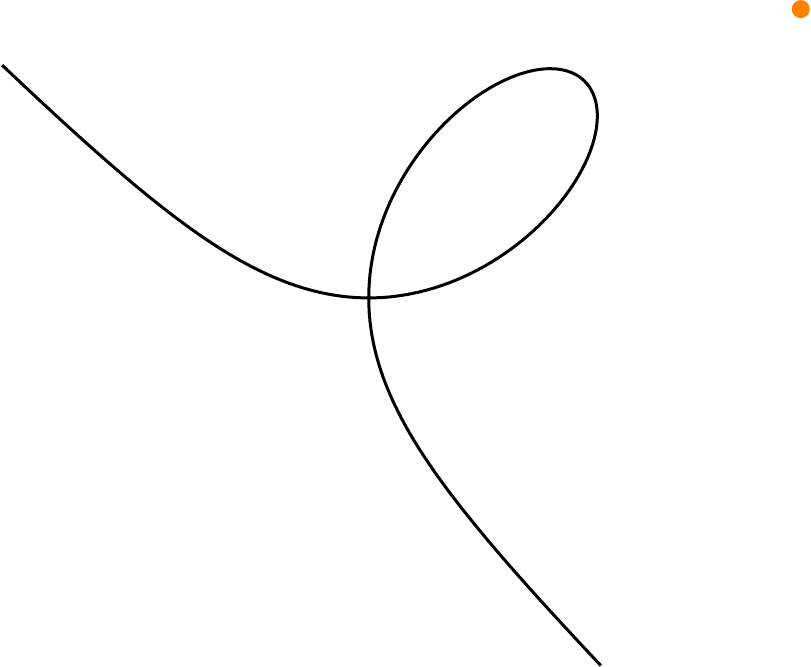}};
    \node [color = orange] at  (2.2,1.5){$L$};
    \node [color = black] at  (-1.7,1.5){$c$};;
 \end{tikzpicture}
 }}%
        \hfill
	\subfloat{{
  \begin{tikzpicture}
    \node at (0,0){\includegraphics[height=100pt, trim= 0 100 0 50, clip]{{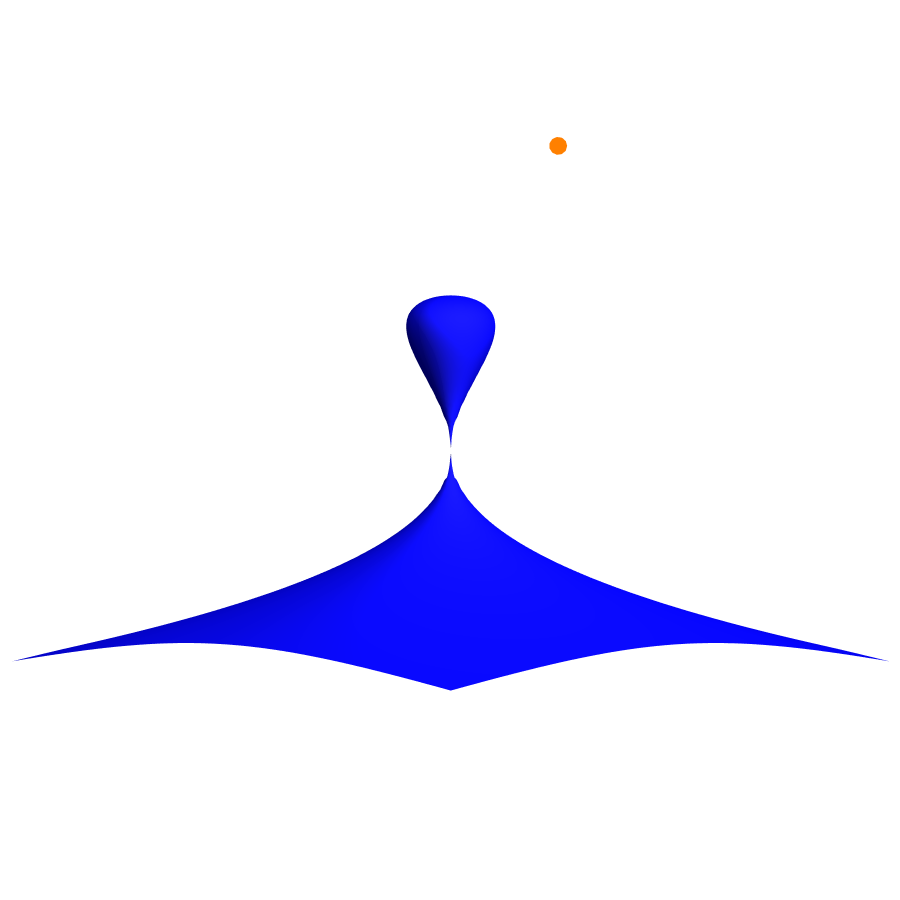}}};
     \node [color = orange] at  (0.8,1.3){$L$};
    \node [color = blue] at  (-1.5,-1.1){$\mathcal{S}$};;
    \end{tikzpicture}
    }}%
    \caption{An initial setting of (left) a folium of Descartes $c: x^3 + y^3 - 6 xy=0$ and (right) a 5th degree surface $\mathcal{S}: x^2 + y^2 + z^4 (z - 1)=0$, and a point light source $L$.}
    \label{fig:setting}
\end{figure}

For the initial setting, let us have a curve $c$ given by a polynomial $\gamma$ or a surface $\mathcal{S}$ with polynomial $\sigma$, and a point $L$ (light source), which is not a singular point of $c$ or $\mathcal{S}$ in $\mathbb{R}^2$ or $\mathbb{R}^3$, respectively (Figure~\ref{fig:setting}).

\subsection{First polar for finding a terminator}

\begin{figure}[!htb]
\centering

         \begin{tikzpicture}
            \node at (0,0){\includegraphics[width=0.75\linewidth, trim=0 100 0 50, clip]{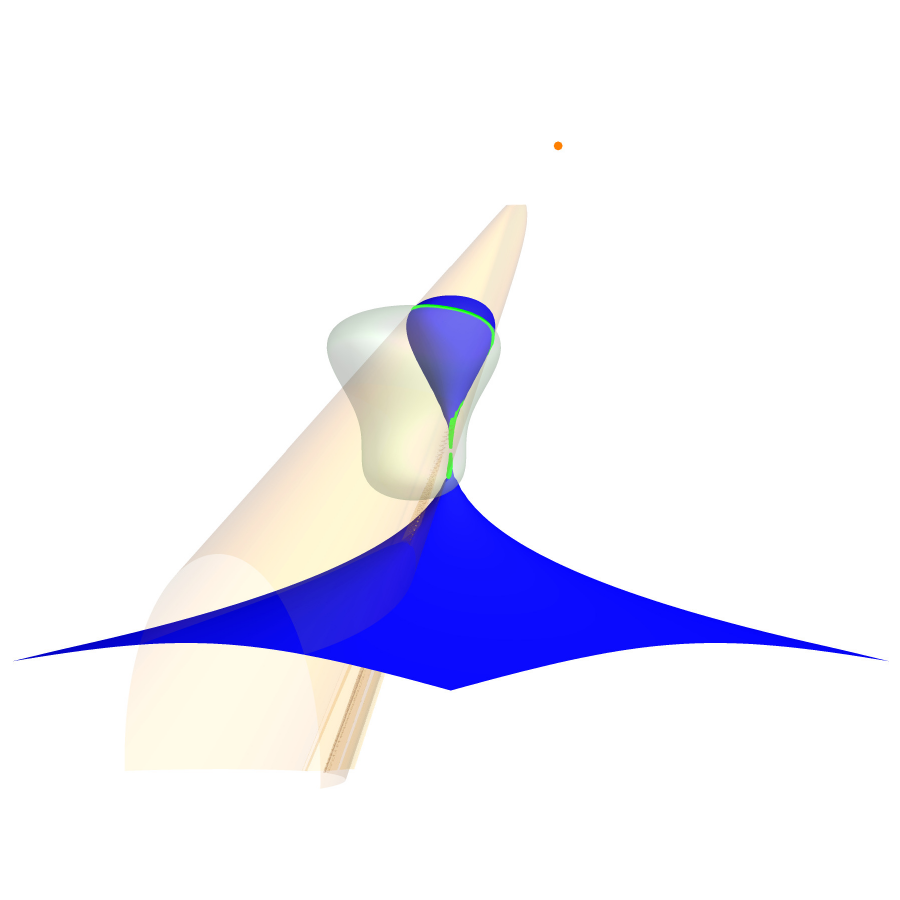}};
            \node at (-6,0){\includegraphics[width=0.3\linewidth, trim= 50 50 100 20, clip]{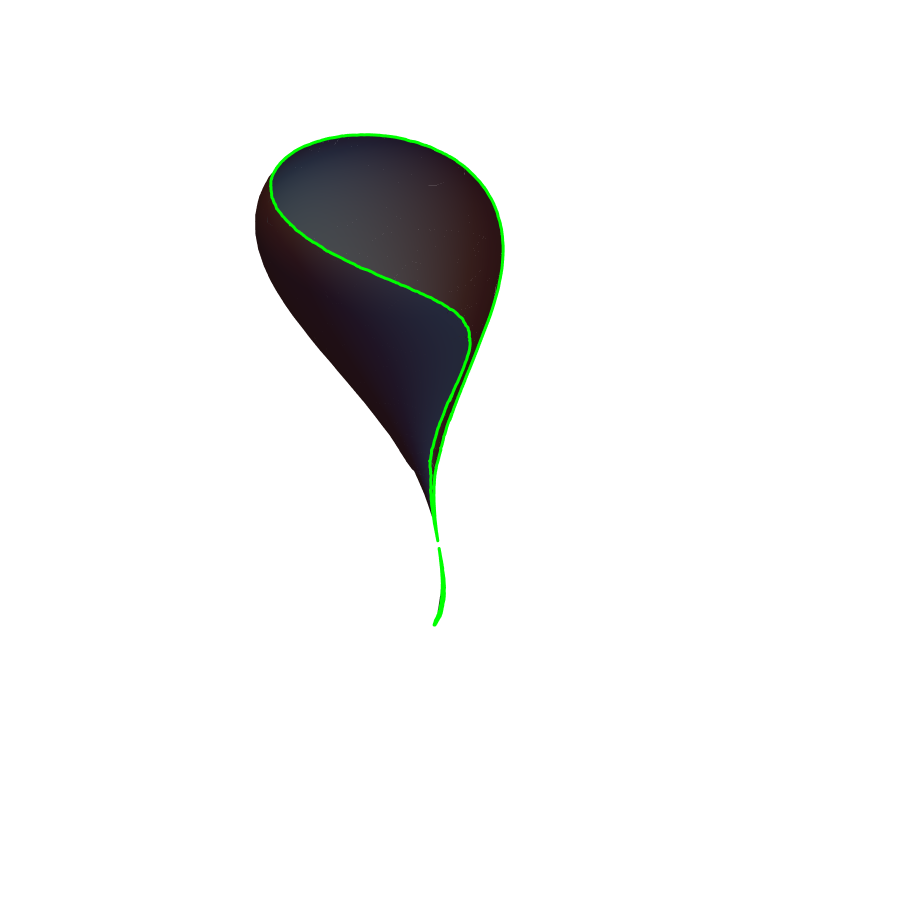} };
            \node [color = orange] at  (1.3,2.7){$L$};
            \node [color = blue] at (1.2,-1.4){$\mathcal{S}$};
            \color{black}\draw[->, >=latex] (-1,2) node[left]{\color{green}$\mathcal{S}_L$}--(-.6,.3);
            \color{black}\draw[->, >=latex] (-2.5,1.3) node[above]{\color{green}$t$}--(0,1);
            \color{black}\draw[->, >=latex] (-2.6,1.3) --(-6,1);
            \color{black}\draw[->, >=latex] (1.2,1.2) node[right]{\color{brown}$\mathcal{T}$}--(0.4,1.6);
           
        \end{tikzpicture}
           
    \caption{The 5th degree surface $\mathcal{S}$ with the first polar surface $\mathcal{S}_L$, terminator $t$, and tangent cone $\mathcal{T}$ with the point light source in $L=[1,0,2]$, and a detail in a different position on the shaded part separated by the first polar and terminator line. The equation of a polar is: $\mathcal{S}_L: 2 x + 3 x^2 + 3 y^2 + z^3 (-8 + 9 z)=0$. The tangent cone is given by a polynomial of degree 10 with 151 terms.}
    \label{fig:terminatorsurface}
\end{figure}
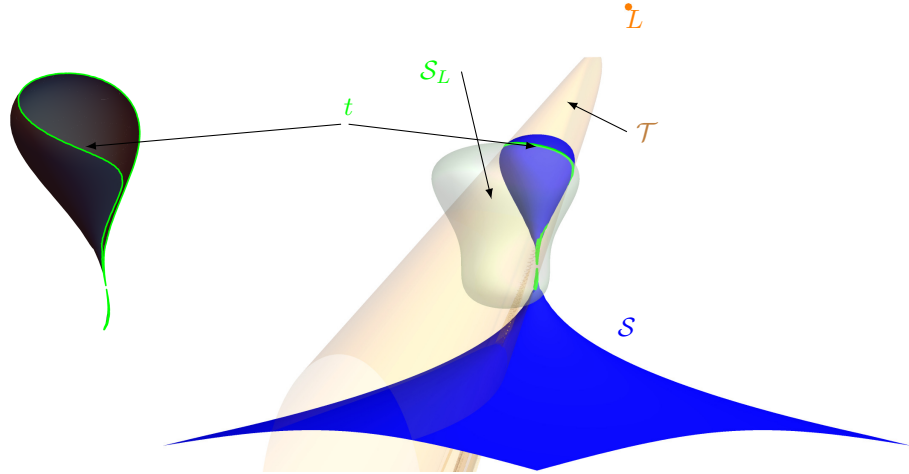

\begin{figure}[!htb]
\centering
    \begin{tikzpicture}      
        \node at (0,0){\includegraphics[width=0.46\linewidth]{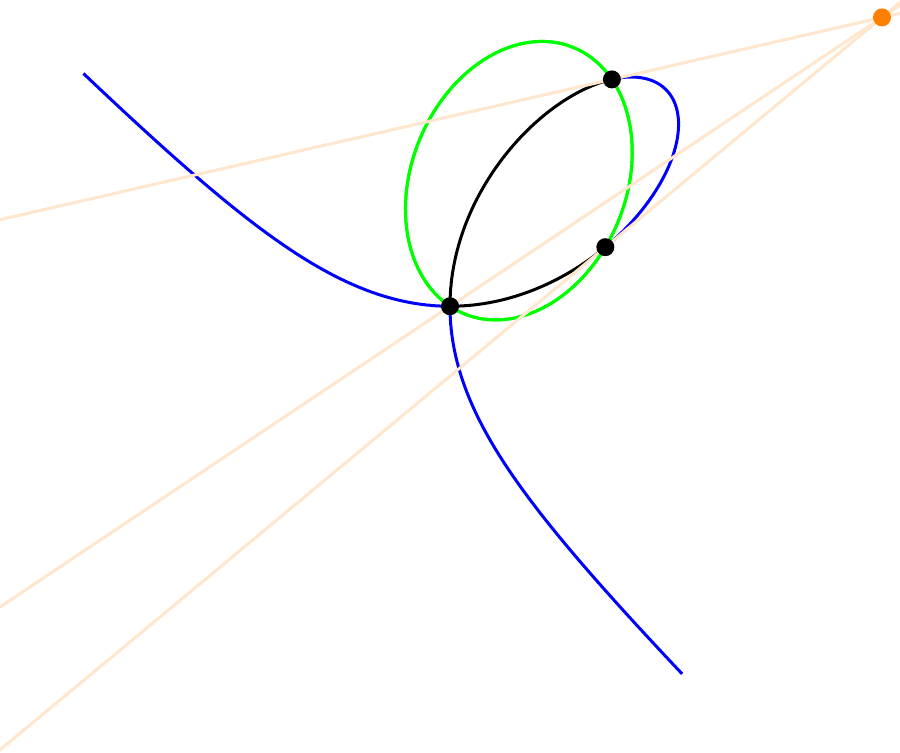}};
        \node [color = orange] at  (3.1,2.3){$L$};
        \node [color = blue] at  (-1.7,1.6){$c$};
        \node [color = green] at  (0,2.4){$c_L$};
        \color{black}\draw[->, >=latex] (3.1,0) node[right]{\color{black}{$c_L\cap c$}}--(0,0.4);
        \color{black}\draw[->, >=latex] (3.1,0) --(1.1,0.85);
        \color{black}\draw[->, >=latex] (3.1,0) --(1.1,2);
    \end{tikzpicture}
    \caption{The folium of Descartes and its first polar $c_L: 2 x^2 - x (6 + y) + y (-4 + 3 y)=0$ (green) with respect to the point light source $L=[6,4]$. Intersections of the curve $c$ and the first polar $c_L$ create the terminator.}
    \label{fig:terminatorcurve}
\end{figure}
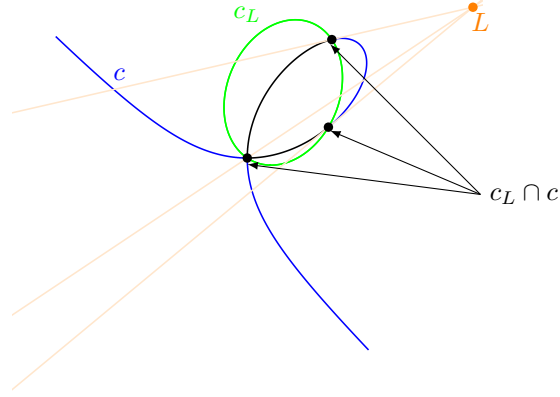

The intersection of the surface $\mathcal{S}$ and its first polar $\mathcal{S}_L$ with respect to a pole $L$ is the set of tangent points of tangents to $\mathcal{S}$ through $L$ (and singular points). The set of these tangent points is the boundary of the self-shade called a terminator $t$ (Figure~\ref{fig:terminatorsurface}), i.e. it is a curve (if exists) that separates the illuminated and shaded parts of a surface. When dealing with curves in a 2-D scene, the intersection is, in the general case, a set of distinct tangent and singular points (Figure~\ref{fig:terminatorcurve}). 

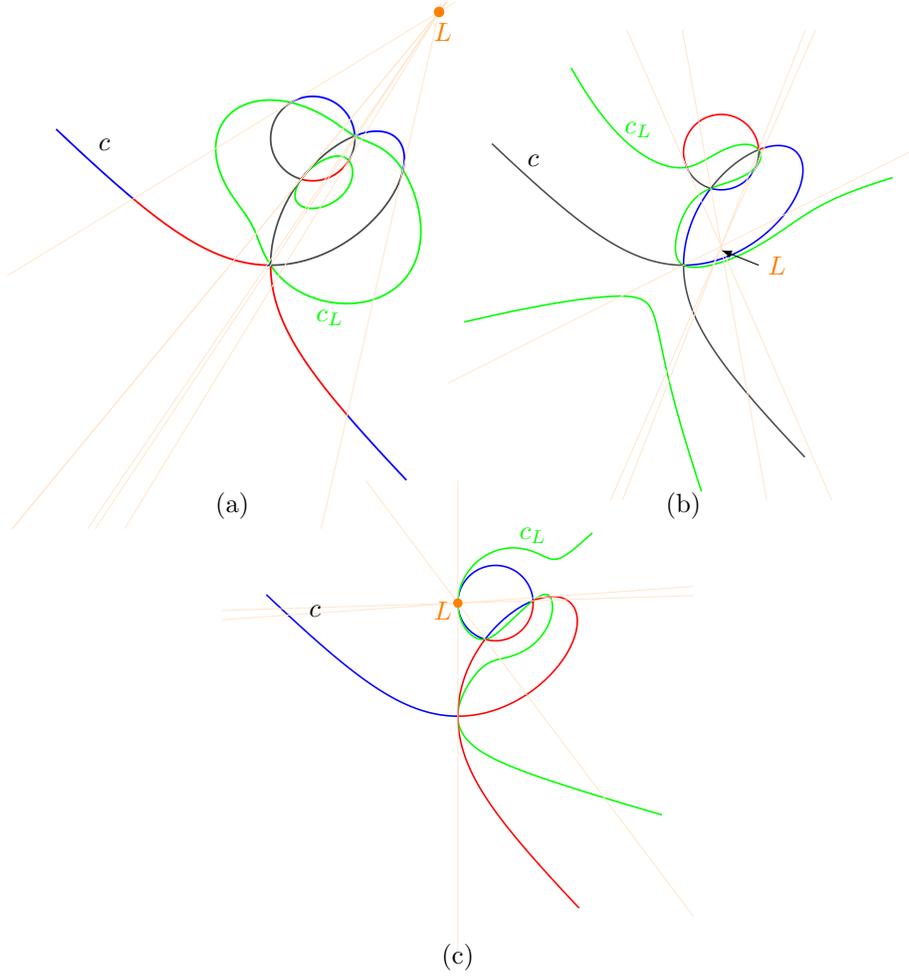
\begin{figure}[!htb]
\centering
        \begin{tikzpicture}
        \node at (-3,0){\includegraphics[width=0.44\linewidth]{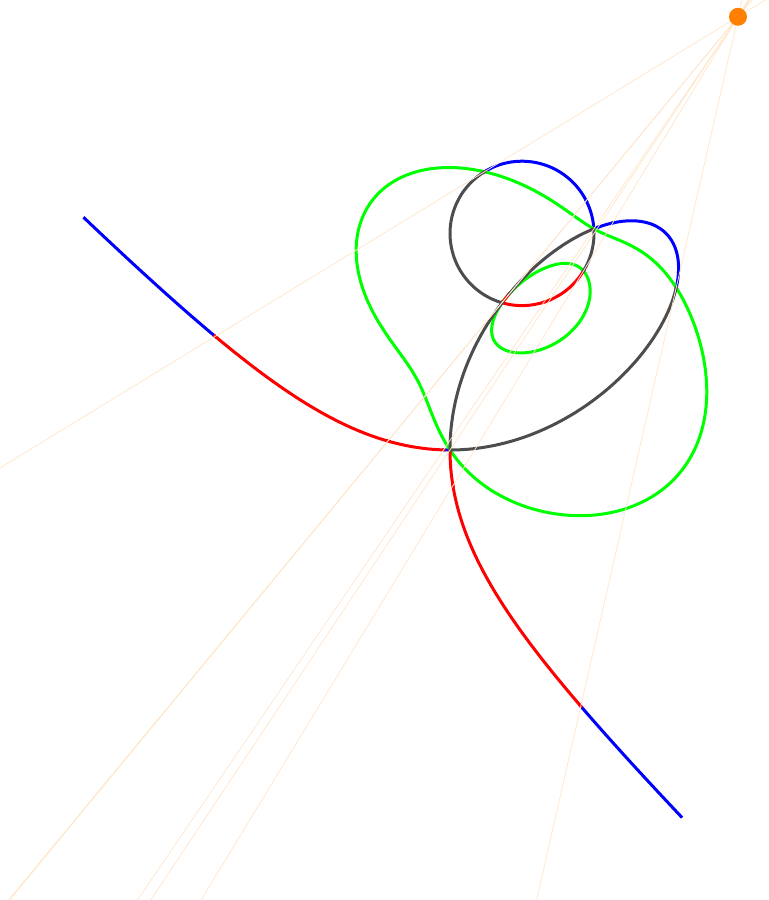} };
        \node at (0,-6){\includegraphics[width=0.46\linewidth]{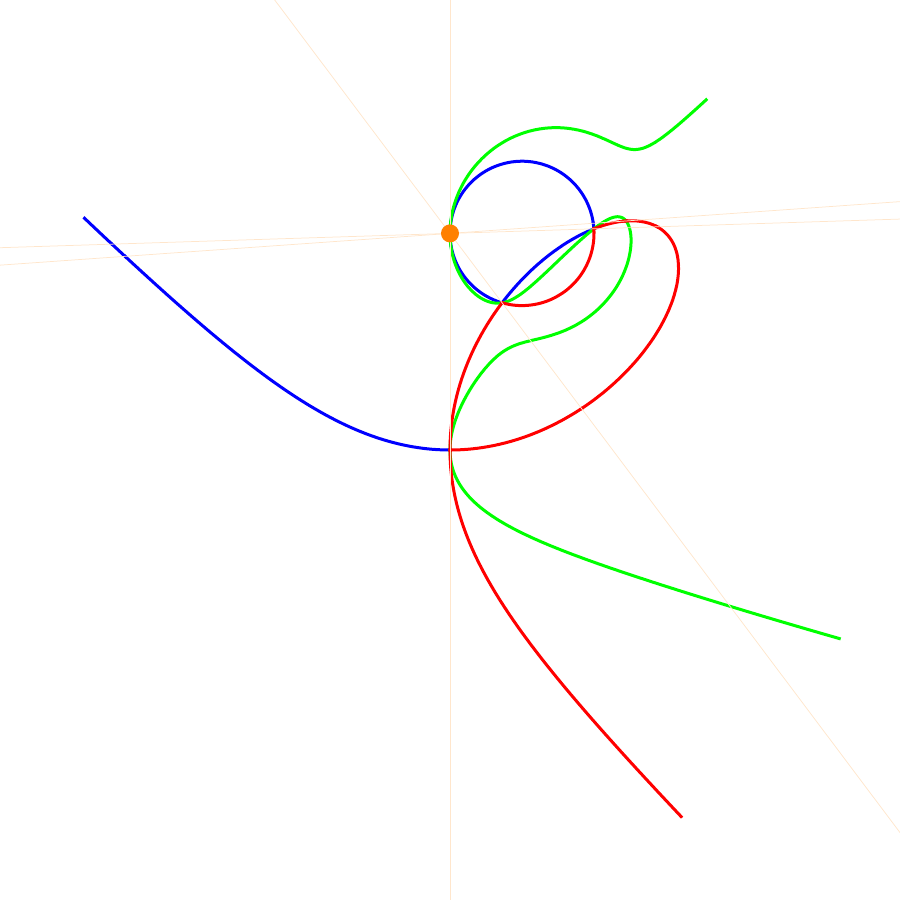}};
	\node at (3,0){\includegraphics[width=0.46\linewidth]{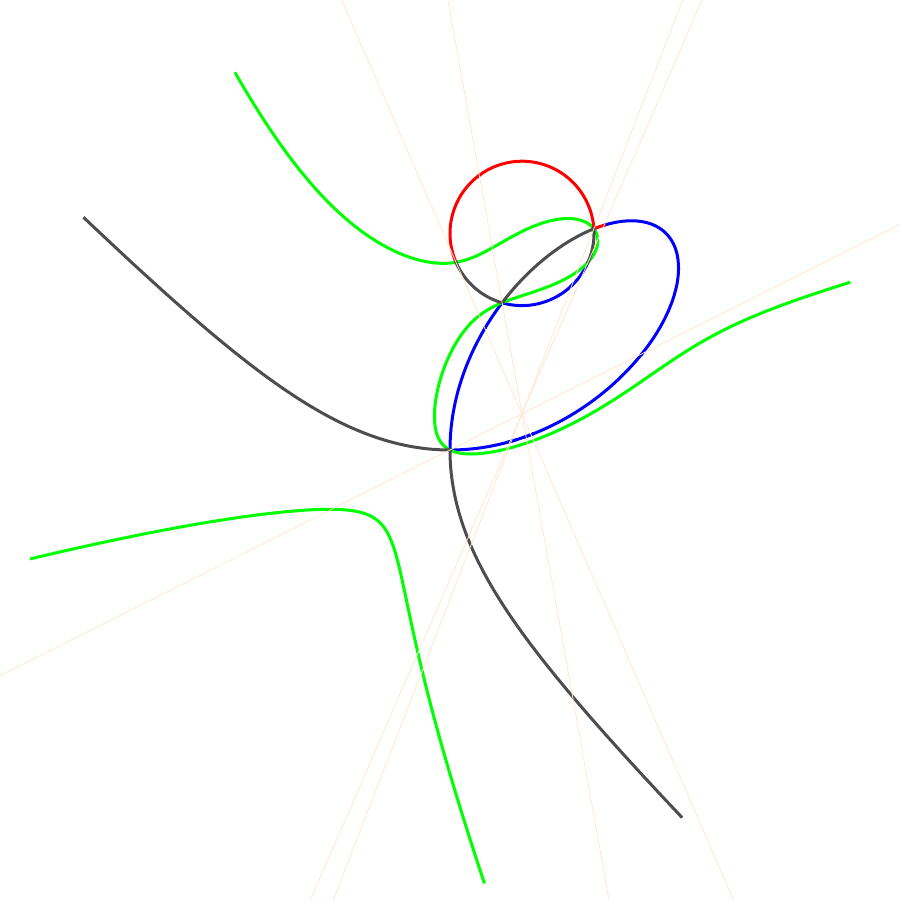} };
        \node [color = orange] at  (-0.2,3.1){$L$};
        \color{black}\draw[->, >=latex] (4,0) node[right]{\color{orange}{$L$}}--(3.5,0.2);
        \node [color = orange] at  (-0.2,-4.6){$L$};
        \node [color = black] at  (-4.7,1.6){$c$};
        \node [color = black] at  (1,1.4){$c$};
        \node [color = black] at  (-1.9,-4.6){$c$};
        \node [color = green] at  (-1.7,-0.7){$c_L$};
        \node [color = green] at  (2.4,1.8){$c_L$};
        \node [color = green] at  (1,-3.6){$c_L$};
        \node [color = black] at  (-3,-3.2){(a)};
        \node [color = black] at  (0,-9.2){(c)};
        \node [color = black] at  (3,-3.2){(b)};
        
         \end{tikzpicture}

    \caption{A polynomial curve $c: \gamma=((x - 1)^2 + (y - 3)^2 - 1) (x^3 + y^3 - 6 xy)$ consisting of a circle and folium of Descartes. The point light source $L$ is located at three different positions: (a) $[4,6]$, (b) $[1,\frac{1}{2}]$, and (c) $[0,3]$ corresponding to $\gamma(L)>0$, $\gamma(L)<0$, and $\gamma(L)=0$, respectively. The blue components are illuminated, the red components are shaded by the blue components, and the black component is separated by the first polar (green).}
    \label{fig:polar}
\end{figure}

The first polar $\mathcal{S}_L$  divides the space into three subspaces: $\sigma_L>0,\ \sigma_L=0,\  \sigma_L<0$. The choice of the subspace that contains the illuminated regions depends on the position of the point $L$ with respect to the surface $\mathcal{S}$, that is, the value of the polynomial $\sigma(L)>0,\ \sigma(L)<0$, or $\sigma(L)=0$ (Figure~\ref{fig:polar}). As a consequence of the definition of polar surface, it holds that for $\sigma(L)>0$ or $\sigma(L)<0$ we have $\sigma_L(L)>0$ or $\sigma_L(L)<0$, respectively. In other words, if $L$ lies in the subspace given by $\sigma>0$, then it also lies in the subspace given by $\sigma_L>0$ and vice versa. The illuminated regions of $\mathcal{S}$ lie in the same subspace as the light source $L$. If $L$ lies on $\mathcal{S}$, then $\sigma(L)=0$ and $L$ also lie on its first polar $\mathcal{S}_L: \sigma_L(L)=0$, and both subspaces are illuminated. 

The subspace separated by the first polar, which is not illuminated, is self-shaded by $\mathcal{S}$ (we refer to this component as polar separated). 

In particular, when dealing with quadratic surfaces, the first polar is linear and provides constraints for separating illuminated and shaded regions on the surface. However, in the case of higher-degree ($n>2$) surfaces, some regions may still shade other parts of the illuminated subspace separated by the first polar. 

\subsection{Tangent cones}

To find the boundary of the shadow cast by $\mathcal{S}$ illuminated from $L$ on itself or on other surfaces, we can use the projection of the terminator $t$ from $L$. We construct a tangent conical surface\footnote{For the sake of simplicity, we further abbreviate to a tangent cone.} $\mathcal{T}$ with terminator $t$ as a generator and vertex in the point light source $L$. Let $Q$ be an arbitrary point on the terminator $t$. The tangent cone $\mathcal{T}$ is the set of lines $LQ$ for all points $Q$. 
        The line $LQ$ is the set of points $X\in\mathbb{R}^3$ that satisfy the following equations with a parameter $a\in \mathbb{R}$
        \begin{equation}
        aL+(1-a)Q=X.
        \end{equation}
The implicit equation of the tangent cone $\mathcal{T}$ is the solution of the polynomial system:
        \begin{equation}
        \label{eq:cone}
        \mathcal{T}: \sigma(Q)=0 \wedge \sigma_L(Q)=0 \wedge aL+(1-a)Q-X=0.
        \end{equation}
 Computing the Gr{\" o}bner basis of the system
    \begin{displaymath}       
        \{\sigma(Q),\ \sigma_L(Q),\ aL+(1-a)Q-X\}   
    \end{displaymath}
        and eliminating the coordinates $q_1,q_2,q_3$ of $Q$ and the parameter $a$, we obtain the polynomial $\theta$ in variables $x,y,z$ representing the tangent cone
     \begin{equation}
        \label{eq:conefin}
         \mathcal{T}: \theta=0.
     \end{equation}

Similarly, in the 2-D case, a tangent cone becomes a pencil of tangent lines through $L$ and the intersections of the given curve and its first polar. The computation of the equations of the tangent lines is similar. The polynomial equation of the pencil of tangents is given by their product.  

\subsection{Decomposition into subcones and subregions}

\begin{figure}[!htb]
\centering
    \setlength{\lineskip}{-7pt}
    \subfloat{{\includegraphics[width=.24\textwidth]{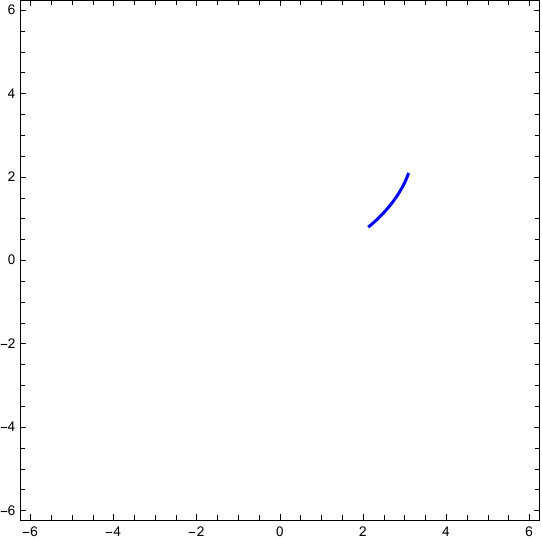} }}
    \subfloat{{\includegraphics[width=.24\textwidth]{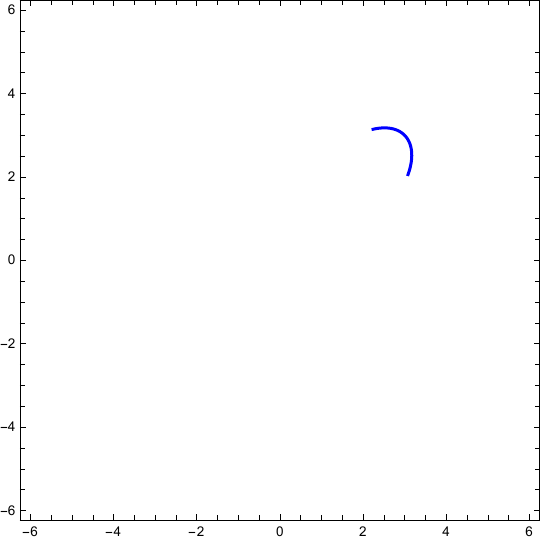} }} 
    \subfloat{{\includegraphics[width=.24\textwidth]{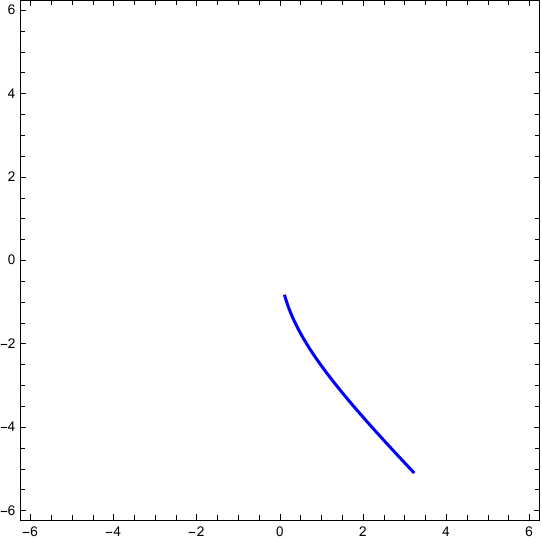} }}
    \subfloat{{\includegraphics[width=.24\textwidth]{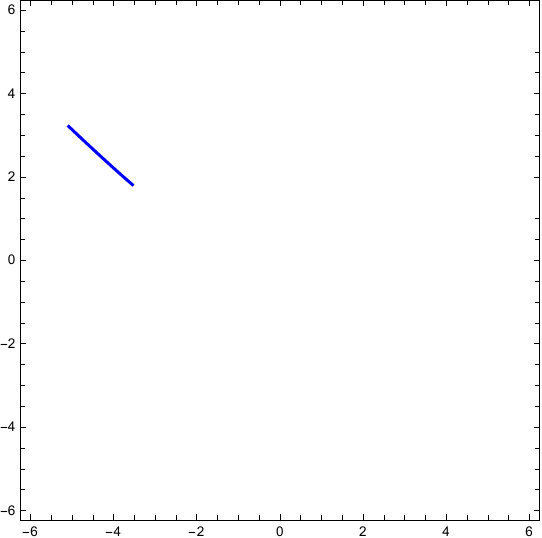} }}
    \newline
    \subfloat{{\includegraphics[width=.24\textwidth]{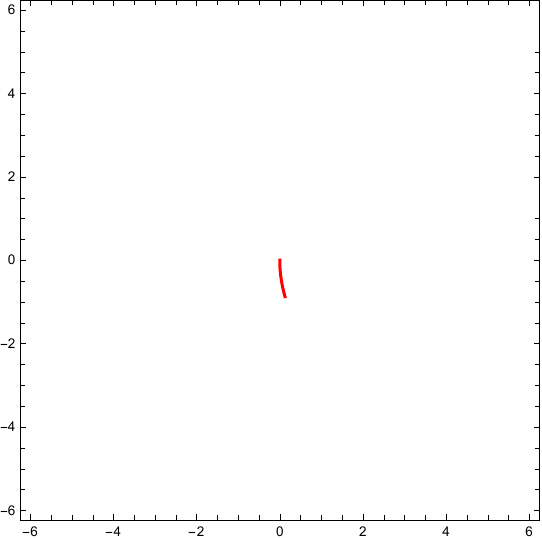} }}
    \subfloat{{\includegraphics[width=.24\textwidth]{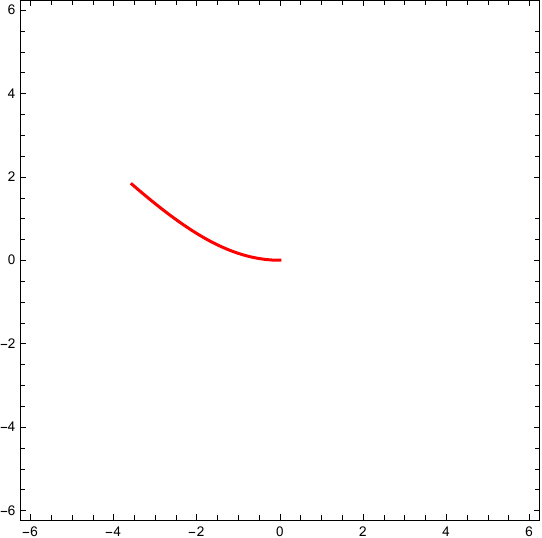} }} 
    \subfloat{{\includegraphics[width=.24\textwidth]{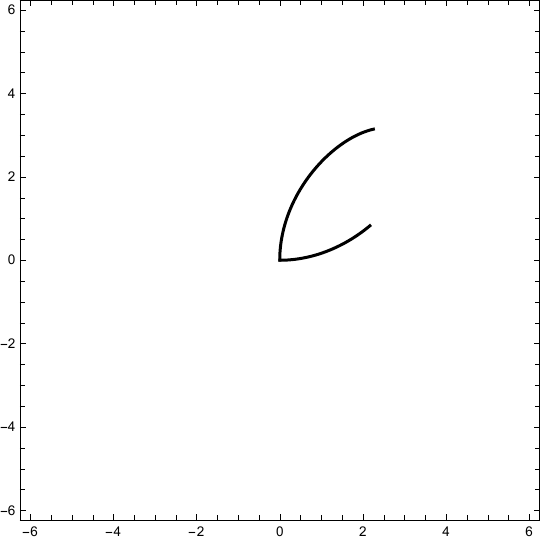} }} \hspace{.24\textwidth} 
    \newline
    \subfloat{{
    \begin{tikzpicture}
        \node at (0,0){\includegraphics[width=0.57\textwidth]{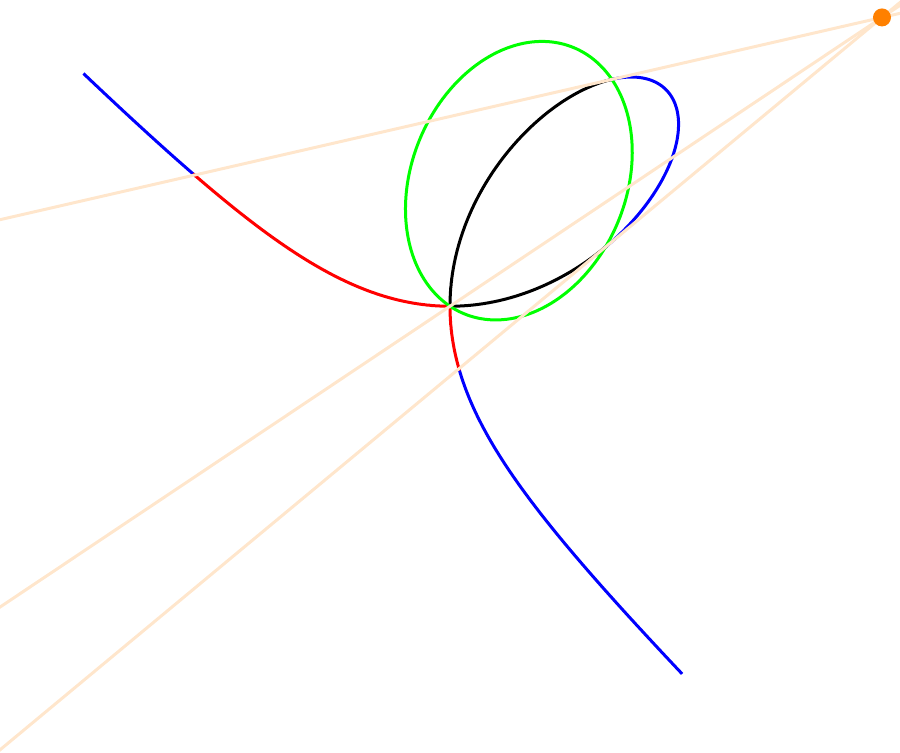}};
        \node [color = orange] at  (3.8,2.8){$L$};
        \node [color = black] at  (-2.1,1.9){$c$};
        \node [color = green] at  (-0.2,2.6){$c_L$};
        \color{black}\draw[->, >=latex] (3.4,0) node[right]{\color{blue}{illuminated}}--(1.6,-1.8);
        \color{black}\draw[->, >=latex](3.4,0) --(2,2);
        \color{black}\draw[->, >=latex](3.4,0) --(-2.5,2.3);
         \color{black}\draw[->, >=latex] (-2.1,0) node[left]{\color{red}{self-shaded}}--(0,0.2);
        \color{black}\draw[->, >=latex](-2.1,0) --(-1,.8);
        \color{black}\draw[->, >=latex] (0,-2) node[left]{\color{black}{polar separated}}--(0.1,1.1);
        \color{black}\draw[->, >=latex](0,-2) --(0.4,.6);
        \end{tikzpicture}
    }}

    \caption{A catalog of components: illuminated ({\color{blue}{blue}}), self-shaded ({\color{red}{red}}), separated by the first polar ({\color{black}{black}});  and their composition.}
    \label{fig:catalog}
\end{figure}
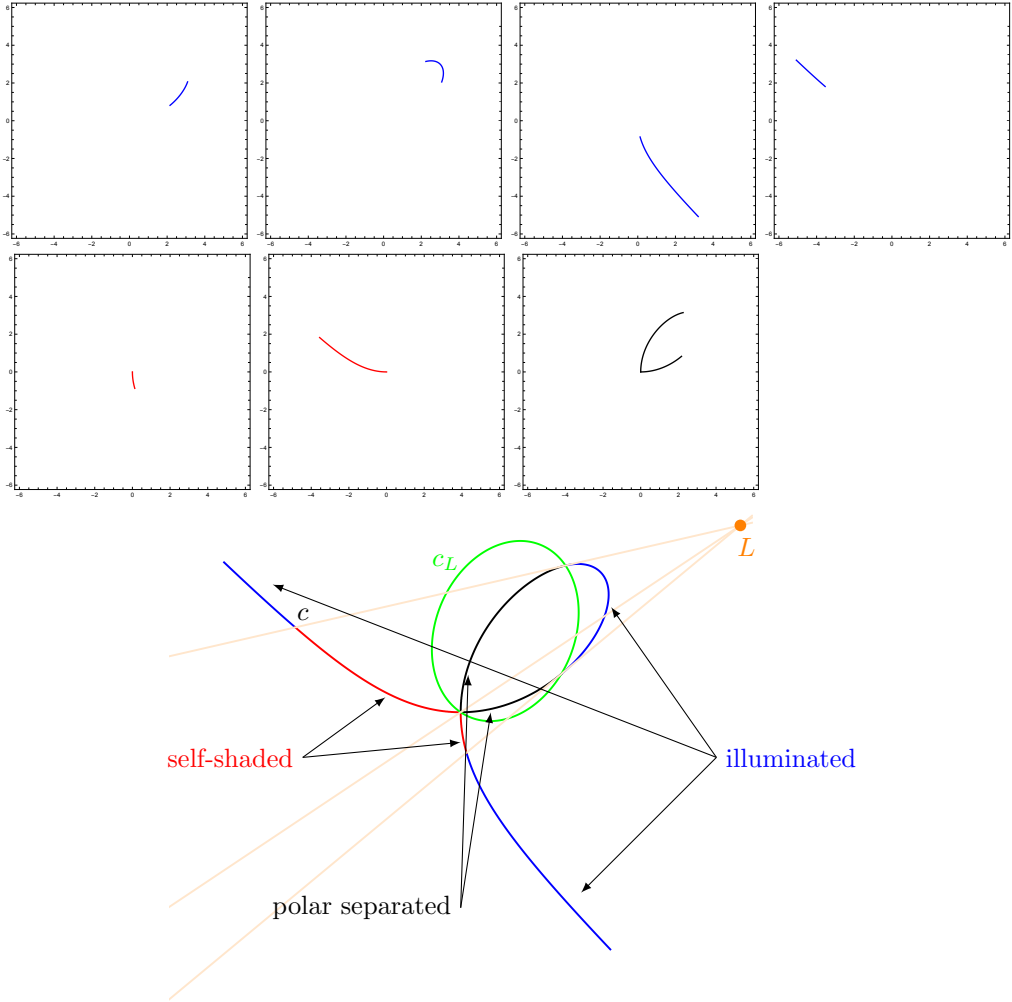

In the next step, we partition the region of the surface selected by the first polar into subregions (Figure~\ref{fig:catalog}). The aim is to separate all illuminated and self-shaded parts (in the illuminated subspace separated by the first polar). The tangent cone (in the sense of surface) divides the embedding space $\mathbb{R}^3$ into conical subspaces (subcones). If a subcone contains one component of the surface $\mathcal{S}$, then this component is illuminated. If a subcone contains more components, then one casts shade on the others. Finally, if a subcone is empty, we can omit it from further computations. 

The intersections of the subcones with the surface $\mathcal{S}$ create subregions. We perform CAD to distinguish them in a semi-algebraic representation. The input conditions are $\sigma=0,\  \theta \neq 0$, and one of $\sigma_L>0,\ \sigma_L<0$ or $\sigma_L=0$, depending on the position of $L$ with respect to $\mathcal{S}$. After the computation, each subregion becomes a system of polynomial conditions.   

\subsection{Selection of the nearest subregion}

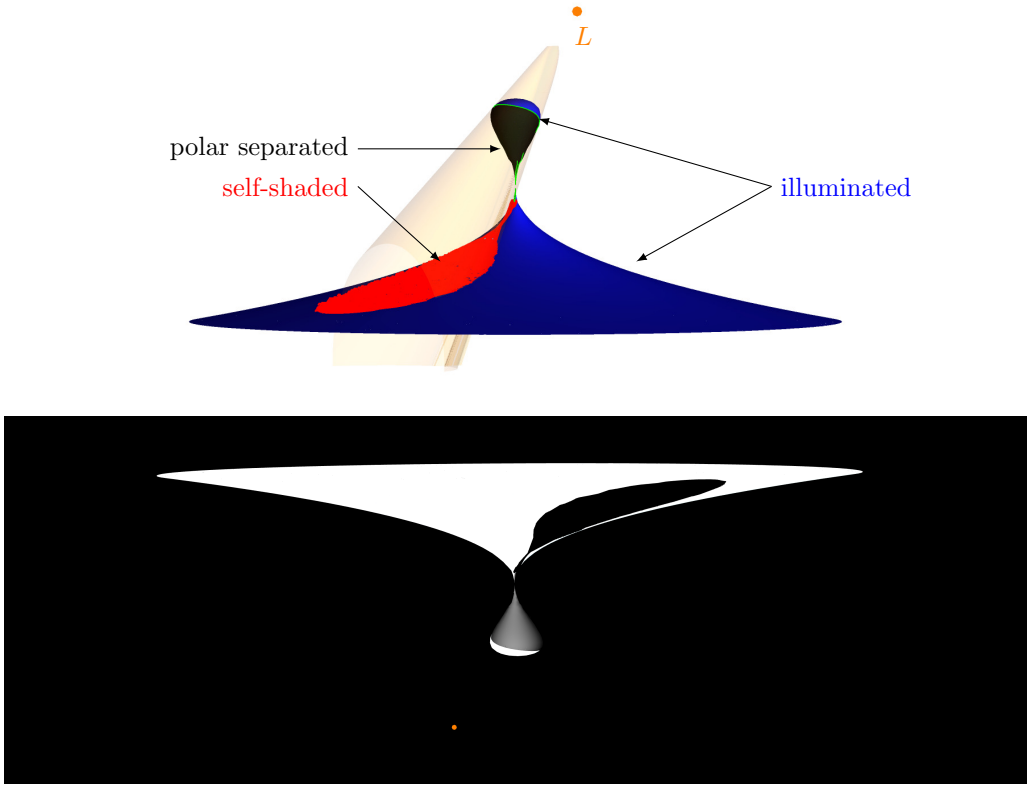
\begin{figure}[!htb]
\centering
\begin{tikzpicture}
    \node at (0,2.5){\includegraphics[width=\linewidth, trim=0 100 0 100, clip]{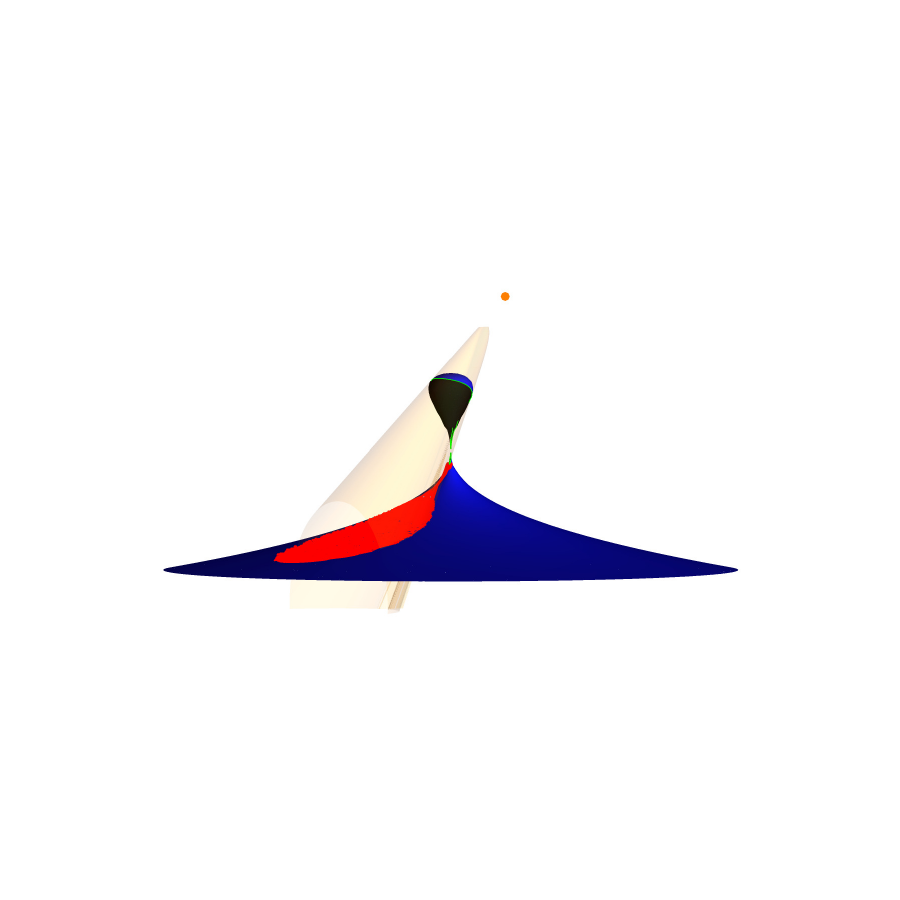}};
    \node [color = orange] at  (.9,4.5){$L$};
    \color{black}\draw[->, >=latex] (3.4,2.5) node[right]{\color{blue}{illuminated}}--(1.6,1.5);
        \color{black}\draw[->, >=latex](3.4,2.5) --(0.3,3.4);
        \color{black}\draw[->, >=latex] (-2.1,2.5) node[left]{\color{red}{self-shaded}}--(-1,1.5);
        \color{black}\draw[->, >=latex] (-2.1,3) node[left]{\color{black}{polar separated}}--(-0.2,3);
    \node at (0,-3){\includegraphics[width=\linewidth, trim=0 0 0 0, clip, angle=180, origin=c]{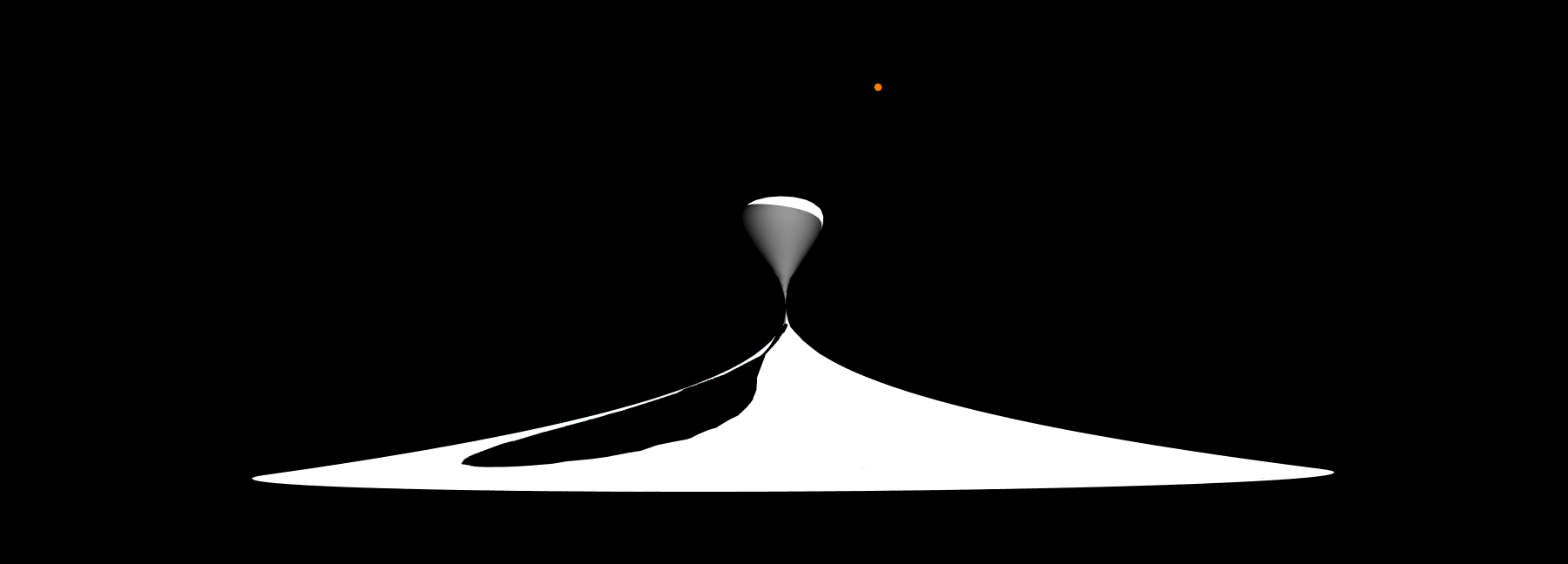}};
\end{tikzpicture}

    \caption{(top) The 5th degree surface with self-shaded regions highlighted in red. CAD divided the surface into 11 regions: 6 illuminated and 5 self-shaded (composed in the figure). (bottom) A slightly edited final visualization.}
    \label{fig:5deg}
\end{figure}

To sort the illuminated and self-shaded subregions, we find their intersections with light rays from the point light source $L$. The subregion with a point on a ray with the smallest distance to $L$ is illuminated, and this region shades the rest. In practice, we choose a random point in each subregion and construct a ray with an initial point $L$. Then, we find existing intersections with all subregions and select the one with minimal distance between the intersection point and $L$. The rest of the subregions with exiting intersections for this ray are automatically in the shaded part, and they are removed from further evaluation for the next rays. See an example of visualization in Figure~\ref{fig:5deg}. 

\subsection{Generalization to more objects}

\begin{figure}[!htb]
\centering
    \begin{tikzpicture}
    \node at (0,0){\includegraphics[width=.65\linewidth]{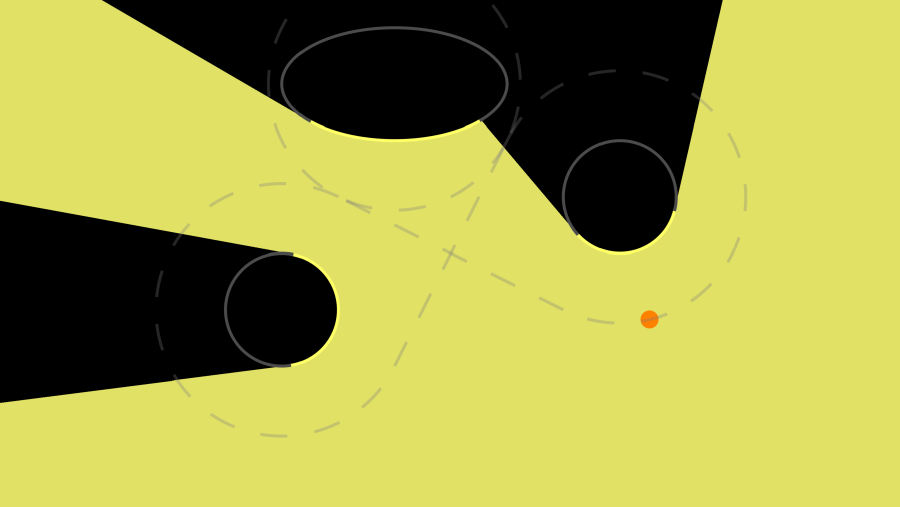}};
    \node [color = orange] at  (2.05,-.9){$L$};
    \color{gray}\draw[->, >=latex](0,-1) --(-.4,-.8);
    \node [color = gray] at (.75,-1){trajectory};
    \node [color = black] at (-.9,-.5){$c_1$};
    \node [color = black] at (1.9,-0.1){$c_2$};
    \node [color = black] at (-.5,.9){$c_3$};
    \end{tikzpicture}

    \caption{A 2-D scene with three conics $c_1: \gamma_1=x^2 + y^2 - 1;~c_2: \gamma_2=(x - 6)^2 + (y - 2)^2 - 1;~c_3: \gamma_3=\frac{(x - 2)^2}{4} + (y - 4)^2 - 1$ illuminated from a point light source $L=[\frac{6527}{1000}, -\frac{173}{1000}]$. The calculation was carried out on the product curve $c$ given by the polynomial $\gamma=\gamma_1\gamma_2\gamma_3$. The animation is available at: \texttt{https://youtu.be/8vLHUG1fUyQ}.}
    \label{fig:circles}
\end{figure}
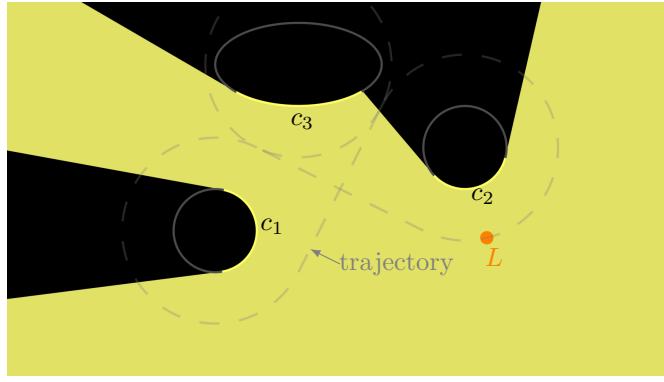

\begin{figure}[!htb]
\centering
\includegraphics[width=.6\linewidth]{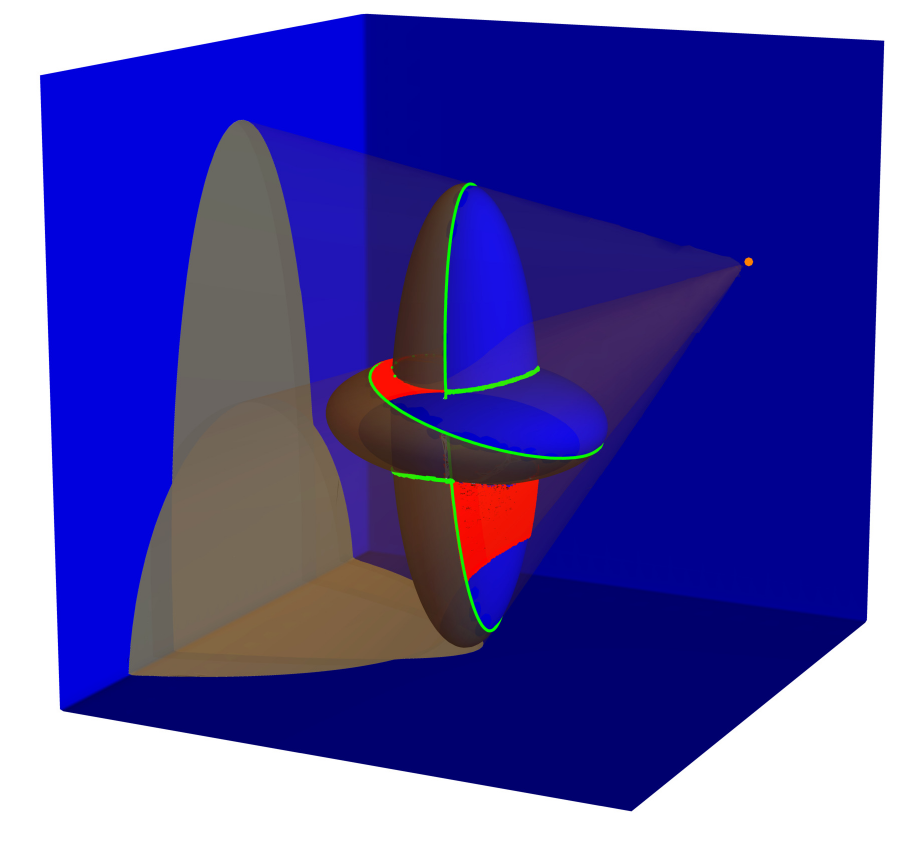}\\%
    \caption{Two intersecting ellipsoids $\mathcal{S}_1: \sigma_1=(\frac{x^2}{4} + \frac{y^2}{2} + 3 z^2 - 3);~\mathcal{S}_2: \sigma_2=x^2 + y^2 + \frac{z^2}{5} - 3$ and three planes $\mathcal{S}_3: \sigma_3=z+\sqrt{15};~\mathcal{S}_4: \sigma_4=x-6;~\mathcal{S}_5: \sigma_5=y+6$            
 illuminated from a point light source $L=[-7,0,3]$. The calculation was carried out on the surface $\mathcal{S}$ given by the polynomial $\sigma=\sigma_1\sigma_2$, and then the shadow was projected onto the planes.}
    \label{fig:2ellips}
\end{figure}

Finally, let us discuss a scene with more objects casting mutual shadows. In general, we can deal with the entire scene of surfaces $\mathcal{S}_1,\dots,\mathcal{S}_k$ for $k\in \mathbb{N}$ as with one product surface $\mathcal{S}=\prod_{k} S_k$ and perform the algorithm described above. Figure~\ref{fig:circles} shows a 2-D scene with three conics treated as one polynomial curve of their products.\footnote{An animation of this scene with a moving point light source is available at \texttt{https://youtu.be/8vLHUG1fUyQ}.}  In this way, we can also deal with surfaces or curves that mutually intersect (see Figures~\ref{fig:polar} and \ref{fig:2ellips}). However, increasing the degree of polynomial constraints (in this case, the sum of the degrees of surfaces) in CAD substantially increases computational complexity.

\begin{figure}[!htb]
\centering
\includegraphics[width=.6\linewidth]{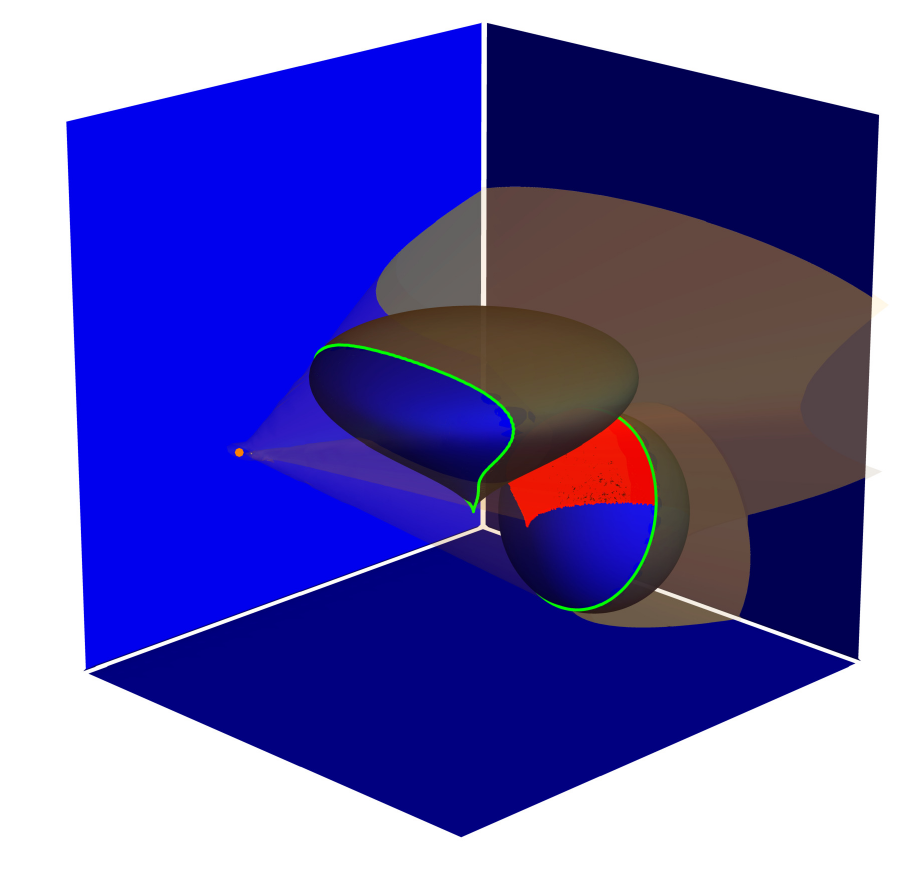}\\%
    \caption{A scene with the surfaces $\mathcal{S}_1: \sigma_1=2 x^2 + 2 y^2 - z^3(4 - z)$ and $\mathcal{S}_2: \sigma_2=(x - 4)^2 + y^2 + (z + 1)^2 - 5$ and three planes $\mathcal{S}_3: \sigma_3=z+\frac{3}{2};~\mathcal{S}_4: \sigma_4=y-6;~\mathcal{S}_5: \sigma_5=x-6$ illuminated from a point light source $L=[-7,0,3]$. The calculation was carried out on the surface $\mathcal{S}$ given by the polynomial $\sigma=\sigma_1\sigma_2$, and then the shadow was projected onto the planes.}
    \label{fig:zeck}
\end{figure}

\begin{figure}[!htb]
\centering
\includegraphics[width=.49\linewidth,trim= 20 100 100 80, clip]{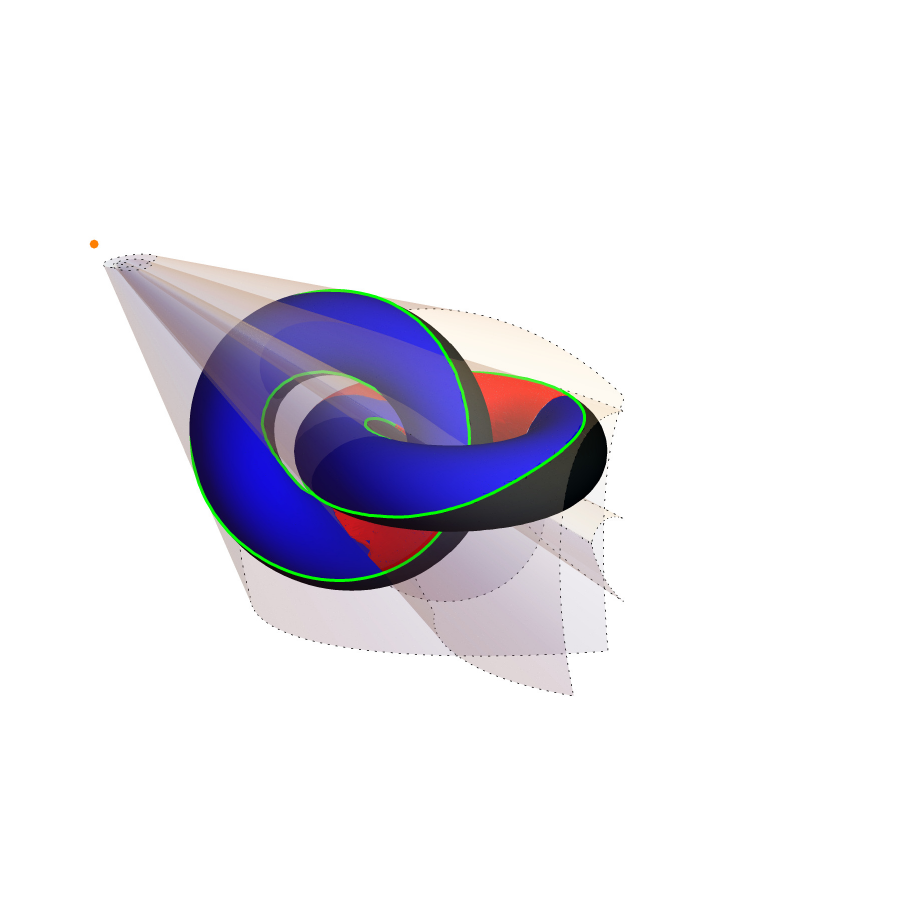}\hfill \includegraphics[trim={0 1cm 0 2cm},clip, width=.49\linewidth]{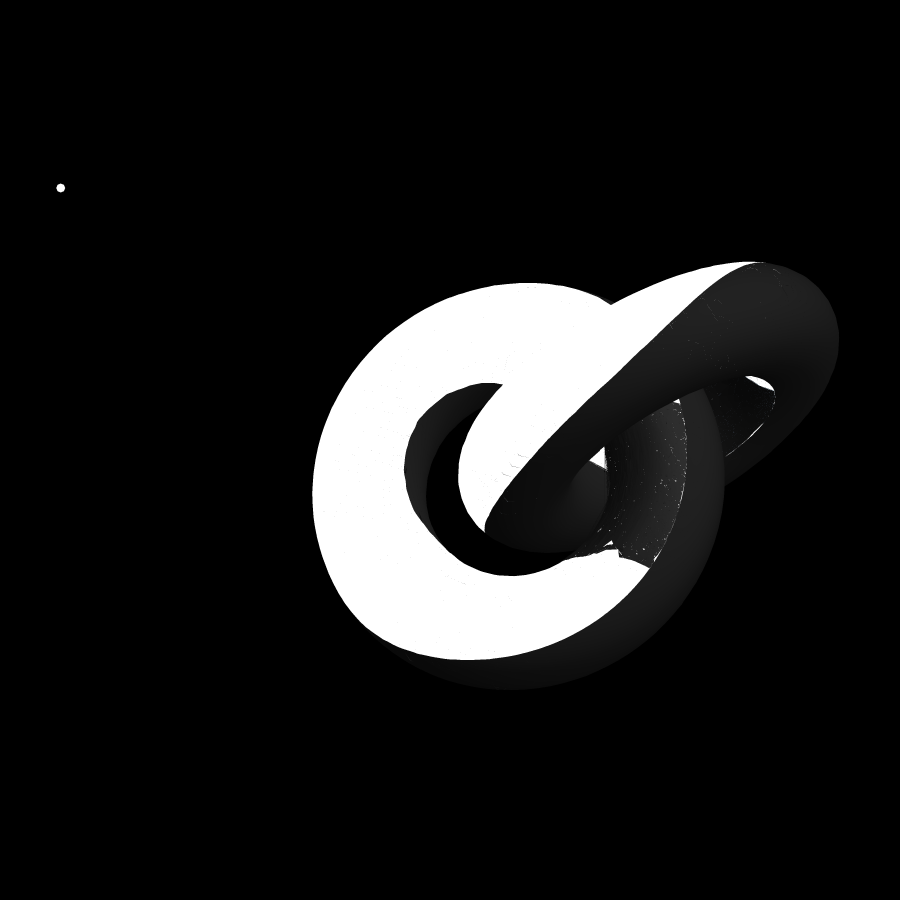} \\%
    \caption{Two entangled, but not intersecting tori $\mathcal{S}_1: \sigma_1=(x^2 + y^2 + z^2 + R^2 - r^2)^2 - 4 R^2 (x^2 + y^2);~\mathcal{S}_2: \sigma_2=((x - 2)^2 + y^2 + z^2 + R^2 - r^2)^2 - 4 R^2 ((x - 2)^2 + z^2)$ for $R=2,~r=\frac{3}{4}$ illuminated from a point light source $L=[5,5,5]$.}
    \label{fig:tori}
\end{figure}

It is often more convenient to have more information about the scene. If we know the order of the surfaces with respect to the light source (considering light rays through each point of the surfaces), we can compute self-shaded parts separately for each surface. Next, we find only the shadows cast by the surfaces closer to those further away. The modification was used in Figures~\ref{fig:2ellips} and~\ref{fig:zeck} to construct shadows cast by surfaces on planes. If the surfaces do not intersect with each other (Figure~\ref{fig:tori}), we can lighten the computation and perform CAD for each surface $\mathcal{S}_1,\dots,\mathcal{S}_k$ separately, keeping the constraint for the common tangent cone $\mathcal{T}$ and the first polar $\mathcal{S}_L$ of $\mathcal{S}$. 

\section{Discussion and further issues}
Let us return to several critical places that appear throughout the construction. The most crucial factor in considering the described method is the computational complexity. Both computational algorithms: the Gr{\" o}bner basis and cylindrical algebraic decomposition depend on the number of constraints and the degree of polynomials (and the number of variables). In particular, a polynomial of degree $n$ has the first polar of degree $n-1$, so the terminator has degree $n(n-1)$. To obtain the GB, we have used the Buchberger algorithm (more precisely, the option: {\sl MonomialOrder $\rightarrow$ EliminationOrder, Method $\rightarrow$ "Buchberger"} in Wolfram Mathematica), and we did not encounter many difficulties. However, in our experience, surfaces given by polynomials of degrees higher than or equal to 4 might significantly increase the computation time. The decomposition is also highly influenced by the position of the point light source and objects in the scene. The number of subregions (and conditions for their representation) proceeding to the selection of self-shaded and illuminated parts is reflected in the increased computational complexity for finding intersections of subregions with light rays. 

The presented method provides a separation of illuminated and shaded regions. No further properties of realistic illumination are discussed. We did not consider the sidedness of surfaces; that is, if a region of a surface is illuminated from one side, it appears illuminated from both sides. This leads to problems with visualization of the illumination of nonorientable surfaces. Additionally, in some cases, it would be possible to conveniently use alternative parametric representations to find intersections and for final plotting. At last, it is fair to remark that we provide a precise mathematical construction up to the point of its final visualization by software. However, the value of our solution lies in the representation.

\section{Conclusion}

We have discussed a method for the geometric illumination of implicit surfaces. By geometric, we mean precise construction by generating and intersecting geometric objects without approximation by polygonal meshes. The result of the algorithm provided is a representation of each component using polynomial equations and inequalities. Therefore, the method is suitable for precise mathematical or scientific visualizations. The construction of shadows is also closely related to the construction of occluding contours in a central or parallel projection. Thus, we can apply a similar method. 

\section{Acknowledgments}
Jakub \v{R}ada (Charles University) was supported by the grant: SVV 260580.

%
\printbibliography
\end{document}